\documentclass[english,plb,10pt,final]{elsarticle}

\usepackage{hyphenat}
\usepackage[english]{babel}
\usepackage[utf8]{inputenc}
\usepackage[colorinlistoftodos, color=green!40, prependcaption]{todonotes}
\usepackage{amsthm}
\usepackage{mathtools}
\usepackage{physics}
\usepackage{xcolor}
\usepackage{graphicx}
\usepackage[left=23mm,right=13mm,top=35mm,columnsep=15pt]{geometry} 
\usepackage{adjustbox}
\usepackage{placeins}
\usepackage[T1]{fontenc}
\usepackage{lipsum}
\usepackage{csquotes}

\usepackage[pdftex, pdftitle={Article}, pdfauthor={Author},colorlinks = true,allcolors = blue]{hyperref}
\usepackage{lineno}
\usepackage{mhchem}
\usepackage{amsmath}
\usepackage[T1]{fontenc}
\biboptions{sort&compress}

\newcommand\eHIJING{e\textsc{HIJING}}
\newcommand\BeAGLE{\textsc{B}e\textsc{AGLE}}

\begin{document}
\begin{frontmatter}{}
\title{First Study of the Nuclear Response to Fast Hadrons via Angular Correlations between Pions and Slow Protons in Electron–Nucleus Scattering}

\newcommand*{\ANL}{Argonne National Laboratory, Argonne, Illinois 60439, USA}
\newcommand*{\ANLindex}{1}
\newcommand*{\ASU}{Arizona State University, Tempe, Arizona 85287-1504, USA}
\newcommand*{\ASUindex}{2}
\newcommand*{\CSUDH}{California State University, Dominguez Hills, Carson, California 90747, USA}
\newcommand*{\CSUDHindex}{3}
\newcommand*{\CANISIUS}{Canisius University, Buffalo, New York 14208, USA}
\newcommand*{\CANISIUSindex}{4}
\newcommand*{\CMU}{Carnegie Mellon University, Pittsburgh, Pennsylvania 15213, USA}
\newcommand*{\CMUindex}{5}
\newcommand*{\CUA}{Catholic University of America, Washington, D.C. 20064, USA}
\newcommand*{\CUAindex}{6}
\newcommand*{\SACLAY}{IRFU, CEA, Universit\'{e} Paris-Saclay, F-91191 Gif-sur-Yvette, France}
\newcommand*{\SACLAYindex}{7}
\newcommand*{\CNU}{Christopher Newport University, Newport News, Virginia 23606, USA}
\newcommand*{\CNUindex}{8}
\newcommand*{\UCONN}{University of Connecticut, Storrs, Connecticut 06269, USA}
\newcommand*{\UCONNindex}{9}
\newcommand*{\DUQUESNE}{Duquesne University, 600 Forbes Avenue, Pittsburgh, Pennsylvania 15282, USA}
\newcommand*{\DUQUESNEindex}{10}
\newcommand*{\FU}{Fairfield University, Fairfield, Connecticut 06824, USA}
\newcommand*{\FUindex}{11}
\newcommand*{\FERRARAU}{Universit\`a di Ferrara , 44121 Ferrara, Italy}
\newcommand*{\FERRARAUindex}{12}
\newcommand*{\FIU}{Florida International University, Miami, Florida 33199, USA}
\newcommand*{\FIUindex}{13}
\newcommand*{\FSU}{Florida State University, Tallahassee, Florida 32306, USA}
\newcommand*{\FSUindex}{14}
\newcommand*{\GWUI}{The George Washington University, Washington, D.C. 20052, USA}
\newcommand*{\GWUIindex}{15}
\newcommand*{\GSIFFN}{GSI Helmholtzzentrum fur Schwerionenforschung GmbH, D-64291 Darmstadt, Germany}
\newcommand*{\GSIFFNindex}{16}
\newcommand*{\ISU}{Idaho State University, Pocatello, Idaho 83209, USA}
\newcommand*{\ISUindex}{17}
\newcommand*{\ORSAY}{Universit\'{e} Paris-Saclay, CNRS/IN2P3, IJCLab, 91405 Orsay, France}
\newcommand*{\ORSAYindex}{18}
\newcommand*{\INFNFE}{INFN, Sezione di Ferrara, 44100 Ferrara, Italy}
\newcommand*{\INFNFEindex}{19}
\newcommand*{\INFNGE}{INFN, Sezione di Genova, 16146 Genova, Italy}
\newcommand*{\INFNGEindex}{20}
\newcommand*{\INFNRO}{INFN, Sezione di Roma Tor Vergata, 00133 Rome, Italy}
\newcommand*{\INFNROindex}{21}
\newcommand*{\INFNTUR}{INFN, Sezione di Torino, 10125 Torino, Italy}
\newcommand*{\INFNTURindex}{22}
\newcommand*{\INFNPAV}{INFN, Sezione di Pavia, 27100 Pavia, Italy}
\newcommand*{\INFNPAVindex}{23}
\newcommand*{\JMU}{James Madison University, Harrisonburg, Virginia 22807, USA}
\newcommand*{\JMUindex}{24}
\newcommand*{\KNU}{Kyungpook National University, Daegu 41566, Republic of Korea}
\newcommand*{\KNUindex}{25}
\newcommand*{\LAMAR}{Lamar University, 4400 MLK Blvd, PO Box 10046, Beaumont, Texas 77710, USA}
\newcommand*{\LAMARindex}{26}
\newcommand*{\MIT}{Massachusetts Institute of Technology, Cambridge, Massachusetts  02139-4307, USA}
\newcommand*{\MITindex}{27}
\newcommand*{\MISS}{Mississippi State University, Mississippi State, Mississippi 39762-5167, USA}
\newcommand*{\MISSindex}{28}
\newcommand*{\UNH}{University of New Hampshire, Durham, New Hampshire 03824-3568, USA}
\newcommand*{\UNHindex}{29}
\newcommand*{\NMSU}{New Mexico State University, PO Box 30001, Las Cruces, New Mexico 88003, USA}
\newcommand*{\NMSUindex}{30}
\newcommand*{\NSU}{Norfolk State University, Norfolk, Virginia 23504, USA}
\newcommand*{\NSUindex}{31}
\newcommand*{\OHIOU}{Ohio University, Athens, Ohio  45701, USA}
\newcommand*{\OHIOUindex}{32}
\newcommand*{\ODU}{Old Dominion University, Norfolk, Virginia 23529, USA}
\newcommand*{\ODUindex}{33}
\newcommand*{\JLUGiessen}{II Physikalisches Institut der Universitaet Giessen, 35392 Giessen, Germany}
\newcommand*{\JLUGiessenindex}{34}
\newcommand*{\RPI}{Rensselaer Polytechnic Institute, Troy, New York 12180-3590, USA}
\newcommand*{\RPIindex}{35}
\newcommand*{\URICH}{University of Richmond, Richmond, Virginia 23173, USA}
\newcommand*{\URICHindex}{36}
\newcommand*{\ROMAII}{Universit\`a di Roma Tor Vergata, 00133 Rome, Italy}
\newcommand*{\ROMAIIindex}{37}
\newcommand*{\SDU}{Shandong University, Qingdao, Shandong 266237, China}
\newcommand*{\SDUindex}{38}
\newcommand*{\MSU}{Skobeltsyn Institute of Nuclear Physics, Lomonosov Moscow State University, 119234 Moscow, Russia}
\newcommand*{\MSUindex}{39}
\newcommand*{\SCAROLINA}{University of South Carolina, Columbia, South Carolina 29208, USA}
\newcommand*{\SCAROLINAindex}{40}
\newcommand*{\TEMPLE}{Temple University,  Philadelphia, Pennsylvania 19122, USA}
\newcommand*{\TEMPLEindex}{41}
\newcommand*{\JLAB}{Thomas Jefferson National Accelerator Facility, Newport News, Virginia 23606, USA}
\newcommand*{\JLABindex}{42}
\newcommand*{\ULS}{Universidad de La Serena, Benavente 980, La Serena, Chile}
\newcommand*{\ULSindex}{43}
\newcommand*{\UTFSM}{Universidad T\'{e}cnica Federico Santa Mar\'{i}a, Casilla 110-V Valpara\'{i}so, Chile}
\newcommand*{\UTFSMindex}{44}
\newcommand*{\BRESCIA}{Universit\`{a} degli Studi di Brescia, 25123 Brescia, Italy}

\newcommand*{\UCRindex}{45}
\newcommand*{\UCR}{University of California Riverside, 900 University Ave, Riverside, California 92521, USA}
\newcommand*{\BRESCIAindex}{46}
\newcommand*{\GLASGOW}{University of Glasgow, Glasgow G12 8QQ, United Kingdom}
\newcommand*{\GLASGOWindex}{47}
\newcommand*{\YORK}{University of York, York YO10 5DD, United Kingdom}
\newcommand*{\YORKindex}{48}
\newcommand*{\TELAVIV}{University of Tel Aviv, Tel Aviv 6997801, Israel}
\newcommand*{\TELAVIVindex}{49}
\newcommand*{\VT}{Virginia Tech, Blacksburg, Virginia   24061-0435, USA}
\newcommand*{\VTindex}{50}
\newcommand*{\VIRGINIA}{University of Virginia, Charlottesville, Virginia 22901, USA}
\newcommand*{\VIRGINIAindex}{51}
\newcommand*{\YEREVAN}{Yerevan Physics Institute, 375036 Yerevan, Armenia}
\newcommand*{\YEREVANindex}{52}

\newcommand*{\NOWISU}{Idaho State University, Pocatello, Idaho 83209, USA}
 %%%%%%%%%%%%%%% END OF Latex Macros for institute addresses  %%%%%%%%%%%%%%%%%%%%%%%%% 

\author[toUCR,toFIU]{S.J.~Paul \corref{cor}} 
\author[toUCR,toJLAB]{M.~Arratia}
\author[toUTFSM]{H.~Hakobyan} 
\author[toUTFSM]{W.~Brooks} 
\author[toYORK]{A.~Acar}
\author[toJLAB]{P.~Achenbach}
\author[toORSAY]{J.S.~Alvarado}
\author[toANL]{W.R.~Armstrong}
\author[toJLAB]{N.A.~Baltzell}
\author[toINFNFE]{L.~Barion}
% ...

 \address[toANL]{\ANL} 
% ...
\address[toFIU]{\FIU} 
% ...
\address[toORSAY]{\ORSAY} 
 \address[toINFNFE]{\INFNFE} 
% ...
\address[toJLAB]{\JLAB} 
% ...
\address[toUTFSM]{\UTFSM} 
% ...
\address[toUCR]{\UCR} 
% ...
\address[toYORK]{\YORK} 
% ...

\cortext[cor]{Corresponding author.  \textit{Email address:} sebouh.paul@fiu.edu (S.J.~Paul)}
 
 \fntext[toNOWISU]{Current address: Pocatello, Idaho 83209 }

\author{\textbf{(CLAS Collaboration)}}

%\date{\today} % Leave empty to omit a date

\begin{abstract}
We report on the first measurement of angular correlations between high‑energy pions and slow protons in electron–nucleus ($eA$) scattering, providing a new probe of how a nucleus responds to a fast‑moving quark. The experiment employed the CLAS detector with a 5-GeV electron beam incident on deuterium, carbon, iron, and lead targets. 
For heavier nuclei, the pion-proton correlation function is more spread-out in azimuth than for lighter ones, and this effect is more pronounced in the $\pi p$ channel than in earlier $\pi\pi$ studies.
The proton‑to‑pion yield ratio likewise rises with nuclear mass, although the increase appears to saturate for the heaviest targets. These trends are qualitatively reproduced by state‑of‑the‑art $eA$ event generators, including \textsc{BeAGLE}, eHIJING, and \textsc{GiBUU}, indicating that current descriptions of target fragmentation rest on sound theoretical footing. At the same time, the precision of our data exposes model–dependent discrepancies, delineating a clear path for future improvements in the treatment of cold-nuclear matter effects in $eA$ scattering.

\end{abstract}
\end{frontmatter}
\twocolumn
 \section{Introduction} 
 The production and propagation of hadrons through nuclei remain central topics in quantum chromodynamics (QCD). Key questions include: \textit{``How does the process of hadronization manifest within a nucleus?''}, and \textit{``How are different hadrons produced in a single scattering event correlated with each other?''}~\cite{LRP2023}.  Addressing these questions calls for detailed studies of diverse reaction channels, encompassing multiple hadron flavor combinations in the final state and spanning a broad kinematic range, to probe the different timescales of hadronization.

Despite extensive theoretical and phenomenological work on nucleon production in nuclear deep-inelastic scattering (DIS) \cite{E665:1994aiu,Degtyarenko:1997,BEBCWA59:1989ayi,CiofidegliAtti:2004pv,Palli:2009it,Larionov:2018igy,PhysRevD.100.073010,Chang:2022hkt,RoblesGajardo:2022efe}, no dedicated study has correlated the production of low-energy protons with the kinematics of leading hadrons, whose spectra are known to be modified by cold nuclear matter effects as observed by the HERMES~\cite{HERMES:2000ytc,HERMES:2003icw,Airapetian:2007vu,Airapetian:2009jy,Airapetian:2011jp} and CLAS~\cite{CLAS:2021jhm,CLAS:2022oux,CLAS:2012tlh,kbhz-h4jv} collaborations.
The present work addresses this gap by employing pion-proton correlations over a wide azimuthal and rapidity range, thereby linking studies of current and nuclear fragmentation in DIS off deuterium and heavy nuclei. 

\begin{figure}
    \centering
    \includegraphics[width=\linewidth]{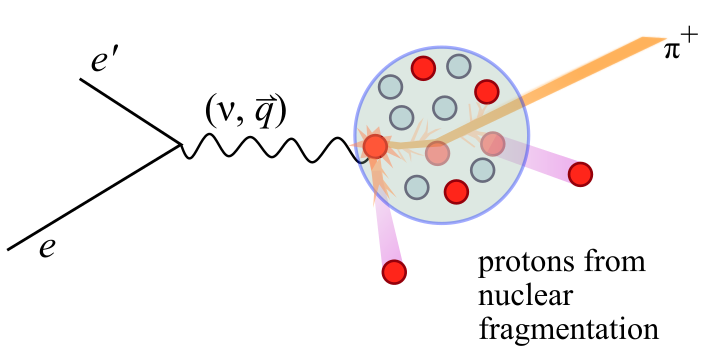}
    \caption{Illustration of nuclear deep‑inelastic scattering featuring a leading charged pion and secondary protons in the final state.}
    \label{fig:illustration}
\end{figure}

In this study, we examine electron–nucleus ($eA$) DIS events that contain a leading $\pi^{+}$ (which retains most of the energy transferred to the struck quark) accompanied by a secondary proton, as illustrated in Fig.~\ref{fig:illustration}. Correlating these two hadrons provides new insight into how the nucleus responds to a fast quark or hadron and how the hadronization process is modified within the nucleus. Most of the observed protons come from fragmentation of the nuclear remnant. 
Because the proton’s rest energy is already present in the initial state, a smaller amount of energy from the fast quark or its hadronic cascade is necessary to knock a proton out of a nucleus than produce a secondary pion with the same momentum.   This allows the current measurement to probe lower energy scales of the hadronization process and transport properties of the nucleus than in existing dipion measurements~\cite{PhysRevC.111.035201}.

Protons produced through nuclear fragmentation, historically referred to as “grey” protons, were investigated in earlier scattering experiments with both hadronic and leptonic beams \cite{E910:1999ozb,E665:1994aiu,Degtyarenko:1997}. 
Measurements of slow protons, together with model calculations, served to estimate the collision centrality, expressed as the mean number of projectile–nucleon interactions.  
Lepton measurements offer a cleaner test of intra-nuclear hadron-cascade models; in hadron collisions, multiple projectile–nucleon interactions complicate particle‐production mechanisms. Similar studies were carried out with neutrino DIS on nuclear targets~\cite{BEBCWA59:1989ayi}, although with much more limited precision.

Ciofi degli Atti~\textit{et al.}~\cite{CiofidegliAtti:2004pv,Palli:2009it} proposed that the dependence of slow-proton production over a wide kinematic range can serve to discriminate among hadronization models, and their calculation reproduced well the data from Ref.~\cite{E665:1994aiu}. Larionov~\textit{et al.}~\cite{Larionov:2018igy} showed that slow-neutrons provide similar insight into intranuclear cascades.  

The study of slow protons (or more generally, slow nucleons) will also play an important role in the future Electron-Ion Collider~\cite{Accardi:2012qut,AbdulKhalek:2021gbh}, since slow nucleons will be measured with dedicated detectors, to enable various studies such as spectator tagging in electron-deuteron collisions, centrality estimations, and hadronization studies in $eA$ collisions~\cite{AbdulKhalek:2021gbh,Chang:2022hkt,RoblesGajardo:2022efe}. 

 Further, the measurement of $\pi p$ correlations in $eA$ DIS can be used as a reference for analogous neutrino measurements, for instance at MINERvA~\cite{ALIAGA2014130}, which overlaps the kinematics of the measurements presented in this work, or with legacy data from NOvA~\cite{osti_935497}.  It was suggested in Ref.~\cite{PhysRevD.100.073010} that left-right asymmetries in $\pi p$ correlations could be measured in $\nu A$ scattering to probe pion-absorption effects.

\section{Experimental setup} 
The data presented here were collected at the Continuous Electron Beam Accelerator Facility (CEBAF) Large Acceptance Spectrometer (CLAS) during the EG2 run period in 2004, with a 5.0-GeV electron beam incident on a
 dual-target system~\cite{Hakobyan:2008kua} consisting of a liquid \ce{^{2}H} target
cell and a \ce{^{}C}, \ce{^{}Fe}, or \ce{^{}Pb} foil target.  The integrated luminosities per target during this run period were 1.0$\times 10^4$ pb$^{-1}$, 4.3$\times 10^2$ pb$^{-1}$, 1.4$\times 10^2$ pb$^{-1}$, and 1.7$\times 10^1$ pb$^{-1}$ for $^2$H, C, Fe, and Pb, respectively.  

The CLAS~\cite{CLAS:2003umf} detector was based on a six-fold symmetric toroidal magnet, which defined six sectors instrumented with drift chambers (DC), time-of-flight scintillation counters (TOF), Cherenkov counters (CC), and an electromagnetic calorimeter (EC). Its polar angular acceptance ranged from 8$^\circ$ to 140$^\circ$ for the drift chambers and TOF, and from 8$^\circ$ to 45$^\circ$ for the Cherenkov counters and calorimeter. The drift chambers functioned in a toroidal magnetic field up to 2~Tm, where the magnet’s polarity made negatively charged particles bend toward the beam pipe. The resulting resolutions for charged particles were $\sigma_p/p=0.5\%$ for the momentum, 2~mrad for the polar angle, and 4~mrad for the azimuthal angle.  The data were selected with a trigger that required at least one electron candidate with momentum $p>500$~MeV$/c$.  

\section{Event selection and observables}  
Following Refs.~\cite{CLAS:2012tlh,CLAS:2021jhm,CLAS:2022asf,PhysRevC.111.035201}, 
electrons were identified by matching negatively charged tracks measured in the DC with hits in the TOF and EC.  Background from $\pi^-$ was suppressed to the $<1\%$ level using the CC and the EC.  Charged pions and protons were identified with TOF hits consistent with those types of hadrons with momentum determined in the DCs.  The selection for $\pi^+$ with momentum above 2.7~GeV$/c$ was further refined with the CC to suppress proton background. Fiducial cuts on momentum and angles were used in order to avoid regions with steeply varying acceptance or low resolution.  The protons' energies were corrected for energy loss in the target using the same procedure as was used in Ref.~\cite{CLAS:2022oux}.

Similar to Refs.~\cite{CLAS:2021jhm,CLAS:2022asf,PhysRevC.111.035201}, we selected events with $Q^{2}>1$~GeV$^{2}/c^2$, $W>2$~GeV/$c^2$ and $2.3<\nu<4.2$~GeV.  Here, $Q^{2}$ is the square of the four-momentum transfer from the electron to the struck nucleon,
\begin{equation}
W=\allowbreak\sqrt{2m_p\nu+m_p^2-Q^{2}}
\end{equation}
is the magnitude of the four-momentum of the virtual photon+struck nucleon system, where $m_p$ is the mass of a proton, $\nu=E-E'$ is the energy transfer, and $E$ and $E'$ are the beam- and scattered-electron energies. The cuts on $Q^2$ and $W$ are conventional for DIS measurements.  The minimum cut on $\nu$ is just above the minimum value of $\nu$ possible with these cuts on $Q^2$ and $W$.   The maximum value of $\nu$ corresponds to a cut on the inelasticity, $y\equiv\nu/E<0.84$.

We selected events with a ``leading'' pion, defined as having fractional energy $z_1=E_h/\nu>0.5$~\cite{HERMES:2005mar,CLAS:2022asf,PhysRevC.111.035201}, where $E_h$ is the energy of the pion.  The purpose of this cut is to increase the likelihood that the pion originated from the struck quark and to avoid ambiguity on which pion in the event is considered ``leading''.  Following Ref.~\cite{Hen:2014nza}, we required the proton to have momentum greater than 350~MeV$/c$, since the proton detection efficiency drops off below this threshold.  The maximum proton momentum in our accepted $\pi p$ event sample was about 2.7 GeV$/c$. In order to ensure that the azimuthal angles of both particles can be reconstructed with reasonable resolution, we further required that both the proton and the leading pion have transverse momenta (defined with respect to the momentum-transfer vector $\vec q$) greater than 70~MeV$/c$.  

We selected events within the range $0.0<\Delta Y<3.0$, where 
$\Delta Y\equiv Y_\pi-Y_p$ is the rapidity separation between the pion and the proton.  Here, $Y_\pi$ and $Y_p$ are the rapidities of the pion and the proton, respectively, each given by
\begin{equation}
Y_h=\frac{1}{2}\log\frac{E_h+p_{z,h}}{E_h-p_{z,h}}, 
\end{equation}
where $E_h$ is the hadron's energy and $p_{z,h}$ is its momentum in the momentum-transfer direction.  This range in $\Delta Y$ contains nearly 99\% of all $\pi^+p$ pairs, and allows for a convenient for binning in $\Delta Y$.  This differs from the $\Delta Y$ range in the di-pion analysis in Ref.~\cite{PhysRevC.111.035201}, where the observed range in $\Delta Y$ for $\pi^+\pi^-$ was approximately $-$0.5 to 2.5.  

We selected particles arising from scattering from either the deuterium or nuclear targets by using the longitudinal vertex position defined by intersecting their trajectories with the beamline. The resulting vertex resolution ensured negligible ambiguity in the target tagging~\cite{CLAS:2021jhm}. 

We used the electron, the leading pion, and the proton variables to measure the 2-dimensional correlation function, defined as
\begin{equation}
    C(\Delta\phi,\Delta Y)=C_0\times \frac{1}{N_{e'\pi}}\left(\frac{d^2N_{e'\pi p}}{d\Delta\phi\,d\Delta Y}\right),
    \label{eq:Cdef}
\end{equation}
where $N_{e'\pi}$ is the number of events with a scattered electron and a leading $\pi^+$, and $\Delta \phi \equiv \phi_\pi - \phi_p$ is the
   azimuthal separation between the pion and proton, as defined around the momentum-transfer
   axis.  We determined the expression in parentheses by taking $N_{e'\pi p}$, the number of events with a scattered electron, a leading $\pi^+$ and a proton in each bin in $|\Delta\phi|$ and $\Delta Y$, and dividing that by the bin widths in both variables ($\pi/8$ and 0.5, respectively), and by a factor of two, since each $|\Delta\phi|$ bin covers two $\Delta\phi$ bins.  These correlation functions are corrected for pair-acceptance effects using Monte Carlo (MC) simulations.  Following Ref.~\cite{PhysRevC.111.035201}, we also used a data-driven method to correct for contamination of the deuterium sample due to scattering off the aluminum endcap walls of the liquid deuterium cells.  Details for both of these corrections can be found in the Supplementary Materials~\cite{supplementary}.

The normalization factor $C_0$ was chosen such that the integral of $C(\Delta\phi,\Delta Y)$ over $\Delta\phi$ and $\Delta Y$ equals unity for deuterium.  We used the same value of $C_0$ for all targets.  Explicitly, the formula we used was 
\begin{equation}
    C_0=\frac{N^D_{e'\pi}}{ N^{D,\rm tot}_{e'\pi p}},
\end{equation}
where $N^D_{e'\pi}$ is the number of events from the deuterium dataset with the scattered electron and leading pion, and $N^{D,\rm tot}_{e'\pi p}$ is the total number of events in the deuterium dataset with the scattered electron, leading pion, and proton, for all kinematic bins.  The purpose of this normalization is to make the correlation function an observable that is insensitive to the absolute scale of the single-proton efficiency of the detector systems.  

From here, we determined the nuclear-to-deuterium ratio of the correlation functions, $R=C_A/C_D$, which can be used to compare the correlation functions of the nuclear targets with those of deuterium.  Since the same $C_0$ value is used for nuclear and deuterium targets, it cancels out in the ratio, which is then equivalent to 
\begin{equation}
    R\equiv\frac{N^A_{e'\pi^+p}/N^A_{e'\pi^+}}{N^D_{e'\pi^+p}/N^D_{e'\pi^+}}.
\end{equation}  

We determined from these correlation functions the azimuthal width $\sigma$, defined by~\cite{PhysRevC.111.035201}
\begin{equation}
    \sigma=\sqrt{\frac{\sum\limits_{i\in \rm bins}C_i\left(\Delta\phi_i-\pi\right)^2}{\sum\limits_{i\in \rm bins}C_i}},
\end{equation}
as well as the azimuthal broadening~\cite{PhysRevC.111.035201} (using carbon as our reference target\footnote{We chose carbon instead of deuterium (as was done in Ref.~\cite{PhysRevC.111.035201}) for the reference target here, since the results for the nuclear targets are more similar to one another than they are to deuterium.}),
\begin{align}
    b=\pm\sqrt{\left|\sigma_A^2-\sigma_C^2\right|},
\end{align}
where the sign of the $\pm$ is that of the expression inside the absolute value.  
This definition of $b$ as a quadrature difference was a natural choice given that the standard deviation of the convolution of two distributions is the quadrature sum of their standard deviations.

\section{Uncertainties} We performed studies on various possible sources of systematic uncertainties using data and simulation studies. For the simulation studies we used the PYTHIA6 event generator and the \textsc{GSIM} package~\cite{GSIM}, which is based on \textsc{GEANT3}~\cite{Brun:1994aa}\footnote{It should be noted that Geant4~\cite{GEANT4:2002zbu} had not been released yet when GSIM was being developed, and that GEANT3 was still the standard at the time that the data were taken.}, to simulate the response of the CLAS detector and dual-target setup~\cite{Hakobyan:2008kua}. The simulation was tuned to provide a reasonable description of the data. 
The possible sources of systematic uncertainties studied include acceptance effects, event selection, and particle misidentification.  For the azimuthal RMS widths and broadenings, there was a small systematic uncertainty due to the finite bin width used to calculate them. Other sources of systematic uncertainty, such as cross-contamination between bins, beam luminosity, trigger efficiency, Coulomb effects,  and time-dependent effects, were found to be negligible.
We show the contributions of various types of systematic uncertainties for the correlation functions and the ratio $R$ in Table~\ref{tab:syst_C} and for the RMS widths and broadenings in Table~\ref{tab:syst_sigma}.  Each of the listed uncertainties are ``point-to-point'', affecting each bin by a different amount (although there may be some correlation between the bins).

 The total relative systematic uncertainty ranged from 15$-$41\% for $C(\Delta\phi,\Delta Y)$, 6$-$21\% for $R$, and 3$-$6\% for $\sigma$.  For $b$, the systematic uncertainty was about 4\% for almost all bins\footnote{The one exception is the highest $\Delta Y$ bin for Fe, where both the statistical and systematic uncertainties are much larger than usual.  The reason for this is that the measured $\sigma$ values for C and Fe are very close together, much less than one statistical error bar.  Therefore, the relative uncertainties on their quadrature difference are very large.}.
Details of these studies on systematic uncertainties can be found in the Supplementary Materials~\cite{supplementary}. 
\begin{table*}[h]
    \centering
    \begin{tabular}{c|c c c c}
        Source & $\Delta C/C$ (D) & $\Delta C/C$ (A) & corr.~D vs A  & $\Delta R/R$\\
        \hline
        Statistics & 1$-$44\% &  2$-$41\% & N  &  2$-$51\% \\
\hline
Endcaps & 0$-$11\% &  -- & -- & 0$-$11\% \\
Particle misid. & 0$-$21\% &  6$-$37\% & Y  &  6$-$21\% \\
Pair acceptance & 15\% &  15\% & Y  &  0\% \\
Event selection & 0$-$7\% &  0$-$7\% & Y &  0\% \\
Bin migration & negligible & negligible & -- & negligible \\
Time dependent effects & negligible & negligible & --  & negligible \\
Luminosity & negligible & negligible & -- & negligible \\
Trigger efficiency & negligible & negligible & --& negligible \\
\hline
Syst. subtotal & 15$-$27\% &  16$-$41\% & -- &  6$-$21\% \\
\hline
Total & 15$-$47\% &  17$-$46\% & -- &  7$-$54\% 
        
    \end{tabular}
    \caption{Summary of statistical and systematic uncertainties on the correlation functions from various sources, listed separately for deuterium (D) and for the nuclear targets (A).  We also list the  uncertainties for the ratio $R=C_A/C_D$, and note whether the systematic uncertainties for the correlation functions are correlated between the nuclear and deuterium targets.}
    \label{tab:syst_C}
\end{table*}

\begin{table*}[]
    \centering
    \begin{tabular}{c|c c c c}
        Source & $\Delta \sigma/\sigma$ (D) & $\Delta \sigma/\sigma$ (A) & corr.~D vs A & $\Delta b/b$ \\
        \hline
        Statistics & 0.4$-$2\% &  0.5$-$4\% & N &  2$-$500\% \\
\hline
Endcaps & $<$1\% &  0\% & N &  0$-$3\% \\
Particle misid. & 2$-$3\% &  2$-$3\% & Y &  1$-$67\% \\
Pair acceptance & 2$-$4\% &  2$-$3\% & Y &  4\% \\
Event selection & 1\% &   1\% & Y &  0\% \\
Finite bin width & 0$-$3\% &  0-1\% & Y &  0$-$4\% \\
Bin migration & negligible & negligible & -- & negligible \\
Time dependent effects & negligible & negligible & -- & negligible \\
Luminosity & negligible & negligible & -- & negligible \\
Trigger efficiency & negligible & negligible & -- & negligible \\
\hline
Syst. subtotal & 3$-$6\% &  3$-$4\% & Y &  4$-$67\% \\
\hline
Total & 3$-$6\% &  3$-$6\% & Y &  4$-$504\% \\

    \end{tabular}
    \caption{Same as Table \ref{tab:syst_C}, for the widths, $\sigma$, and broadenings, $b$.}
    \label{tab:syst_sigma}
\end{table*}

\begin{figure*}
    \centering
    \includegraphics[width=\textwidth]{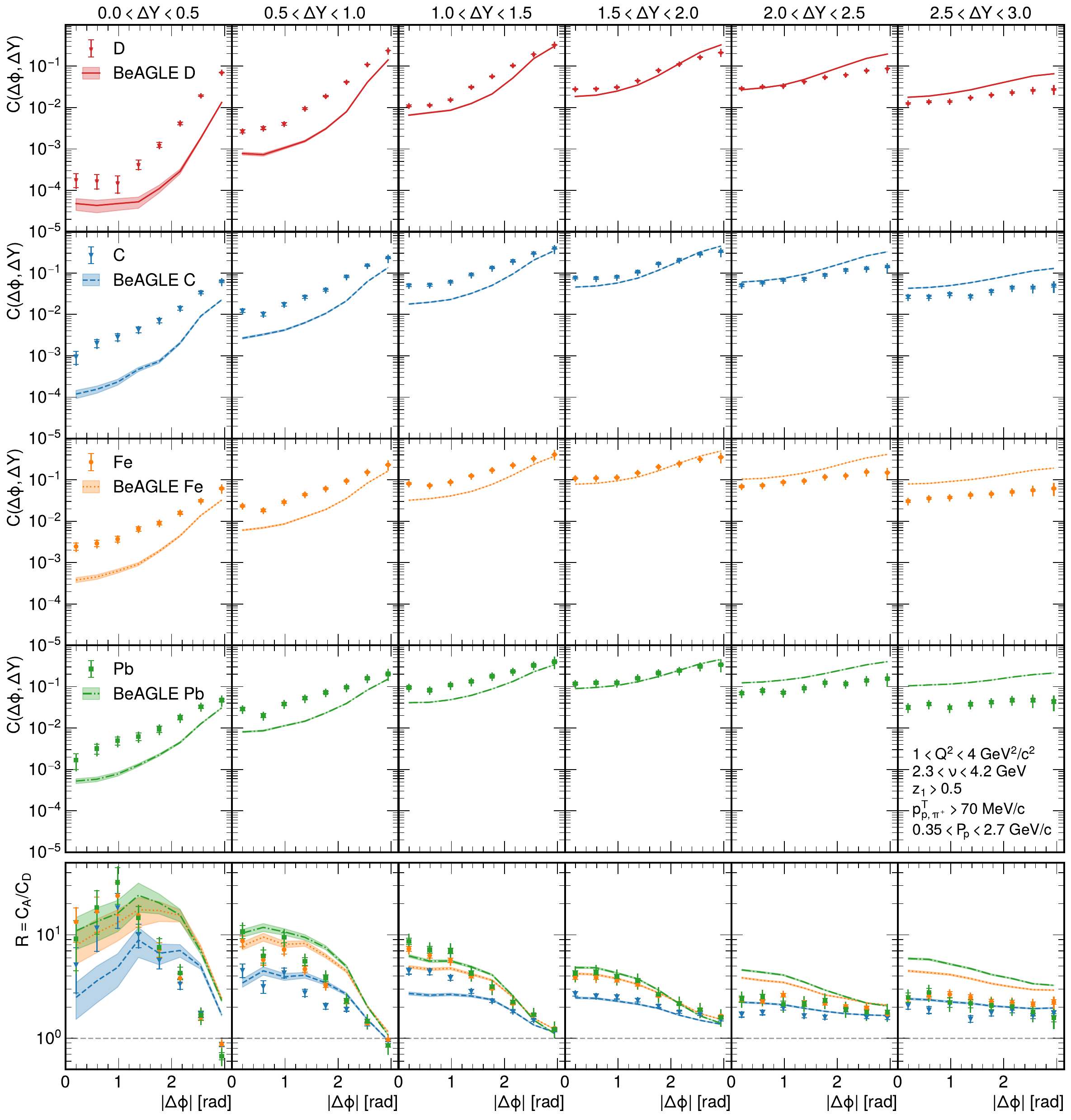}
    \caption{Correlation functions for D, C, Fe, and Pb (from top to bottom, first four rows), with $\Delta\phi$ on the $x$ axis within each panel and different bins of $\Delta Y$ for each column.   These are compared to calculations from the \textsc{BeAGLE} model~\cite{Chang:2022hkt} (curves).  The horizontal caps
in the error bars represent the systematic uncertainties, while the outer bars represent the total systematic and statistical uncertainty (added in quadrature).   Bottom row:  same for the ratios between the nuclear and deuterium correlation functions.}
    \label{fig:corr}
\end{figure*}

 \section{Results and discussion}  
Figure~\ref{fig:corr} shows the correlation function $C(\Delta\phi,\Delta Y)$ with different panels representing different $\Delta Y$ bins and the $x$ axis representing $|\Delta\phi|$\footnote{We bin in $|\Delta\phi|$ rather than $\Delta\phi$ since the observed correlation function is symmetrical with respect to $\Delta\phi=\pi$.}.  We note that the correlation function has a maximum, or ``peak\footnote{Had we shown $\Delta\phi$ from 0 to $2\pi$, as was done in Ref.~\cite{PhysRevC.111.035201}, this would appear as a symmetrical peak, since the correlation function is symmetrical about $\Delta\phi=\pi$.  That is, the correlation function should have the same value at $\Delta\phi=179^\circ$ and at $\Delta\phi=181^\circ$.}'', at $|\Delta\phi|=\pi$ (that is, where the two hadrons are azimuthally opposite one another) in every bin for every target.  
Further, this peak is tallest within the $1.0<\Delta Y<1.5$ bin (third column).  This happens to coincide with the average rapidity separation between the $\gamma^*N$ system and the lab frame (about 1.2), which would be expected if the knocked-out proton rapidity distribution were peaked at that of the lab frame (as would be the case if their initial momentum distribution is isotropic), and the leading pion rapidity distribution were peaked at the $\gamma^*N$ system from which it originated.  This interpretation could be tested by seeing if this observation continues to hold true in future measurements, \textit{e.g} in experiments with the CLAS12 detector~\cite{Burkert:2020akg}, where a larger range in $\nu$ and $Q^2$ allows a larger range of values for the rapidity of the $\gamma^*N$ system.
The peak at $|\Delta\phi|=\pi$ is well pronounced for projections where $\Delta Y$ is small, and more spread out in the slices where $\Delta Y$ is large.

The ratios of the correlation functions for nuclei to those of deuterium are shown in the bottom row of Fig.~\ref{fig:corr}.  The values of these ratios range from near unity to about 30, depending on the target and kinematic bin.  We note that these ratios are smaller for azimuthally opposite pairs ($|\Delta\phi| \approx\pi$) than for azimuthally adjacent pairs ($|\Delta\phi|\approx 0$).  This is similar to observations in dipion-production measurements~\cite{CLAS:2022asf,PhysRevC.111.035201}, which were attributed to absorption effects coupled with geometrical biases caused by survival selection  (that is, hadrons that are absorbed in the nucleus are not detected.)~\cite{Fialkowski:2007pt}.  
 
 We further note that, except when $\Delta Y$ is small and $|\Delta\phi|\approx\pi$, the correlation functions are larger for the nuclear targets than for deuterium (as indicated by the ratio, $R$, being greater than one).  Since the correlation functions are related to the proton yield per leading pion, $N_{e'\pi p}/N_{e'\pi}$, (see Eq.~\ref{eq:Cdef}), it follows that more protons are produced per leading-$\pi^+$ event in nuclei than in deuterium.  This can partially be explained by the fact that there are more protons that can be knocked out of a larger nucleus than a smaller nucleus.  However, this trend appears to saturate, as the values for the correlation functions for iron and lead are within about 30\% of one another despite lead having more than three times as many protons as iron.  In a scenario where only the protons in or near the path of the cascade could be knocked out, the number of protons that could be knocked out would scale with the nuclear radius, which can be proxied by $\sqrt[3]{A}$, rather than the atomic number, $Z$.  In this case, about $\sqrt[3]{A_{\rm Pb}}/\sqrt[3]{A_{\rm Fe}}=\sqrt[3]{208/56}\approx1.55$ times as many protons could be knocked out from Pb than from Fe, which is closer to what we observe than if this were scaled with $Z$.
 Further, the cascade has a limited amount of energy, some of which causes protons to be knocked out and part of which goes into producing other particles, therefore limiting the number of protons that can be knocked out per reaction.  This explanation can be tested by comparing these results to future measurements at higher energy, such as a recent experiment with the CLAS12 detector~\cite{Burkert:2020akg}, which used D, C, Al, Cu, Sn, and Pb targets.  According to this interpretation, the saturation effect should occur at a larger value of $A$ for higher energies than at lower energies.  

 We also note that in the highest $\Delta Y$ bin (rightmost column of Fig.~\ref{fig:corr}), the protons and pions are very weakly correlated with one another, as indicated by the correlation function having little-to-no slope with respect to $\Delta\phi$, and that the average number of protons per pion relative to deuterium (see bottom-right panel of Fig.~\ref{fig:corr}) is $\approx$2 for all three targets.  Our observation that the protons are very weakly correlated with far-forward pions suggests that future measurements could, by selecting events with large or small numbers of protons, seek to enhance or decrease nuclear effects without biasing the forward pion kinematics~\cite{RoblesGajardo:2022efe}.

% comparison to event generator results
We compare our results in Fig.~\ref{fig:corr} with the calculations of the \textsc{BeAGLE} model~\cite{Chang:2022hkt} (version 1.01.03), which was developed as a benchmark model for the Electron-Ion Collider~\cite{Accardi:2012qut,AbdulKhalek:2021gbh}. The \BeAGLE~event generator integrates modules from various sources. The primary interaction is managed by \textsc{PYTHIA6}~\cite{Sjostrand:2006za}. It incorporates nuclear parton distribution functions (PDFs) from \textsc{LHAPDF5}~\cite{Whalley:2005nh}, intranuclear scattering via \textsc{DPMJet}~\cite{Roesler:2000he}, and the geometric density of nucleons through \textsc{PyQM}~\cite{Dupre:2011afa}. Nuclear remnant de-excitations and decays are processed using \textsc{FLUKA}~\cite{Ferrari:2005zk}.  We determined the correlation functions from these simulations by applying Eq.~\ref{eq:Cdef} to the number of simulated events within each kinematic bin.  The value of $C_0$ used for normalizing the simulated correlation functions was determined using the deuterium simulations and is independent of the value used for data.  However, they are close to one another: $C_0^{\rm data}\approx 7.5$ and $C_0^{\rm BeAGLE}\approx 8.1$, indicating that the prediction for the average number of ejected protons per leading-pion event in deuterium is within 10\% of that of the data.  

We find that the trends are similar to those of the data; however, there are numerical discrepancies.  Notably, the predicted distribution in $\Delta Y$ appears to be shifted by one bin, or 0.5 rapidity units compared to the data, causing the peak value of the correlation function to be in the 4th panel ($1.5<\Delta Y<2.0$) instead of the 3rd panel ($1.0<\Delta Y<1.5$) as in the data.  Further, the predicted curves in a given panel appear to match the data points in the panel to the left of it better than those in the same panel.  This could indicate a mis-tuning within the \textsc{BeAGLE} model (possibly in the underlying \textsc{Pythia6} parameters or in \textsc{BeAGLE}'s modeling of Fermi motion or other phenomena), causing either the generated final-state pions to tend towards higher rapidity than those in the data, or the generated protons to tend towards lower rapidity than those in the data.

\begin{figure*}[t!]
    \centering
    \includegraphics[width=\textwidth]{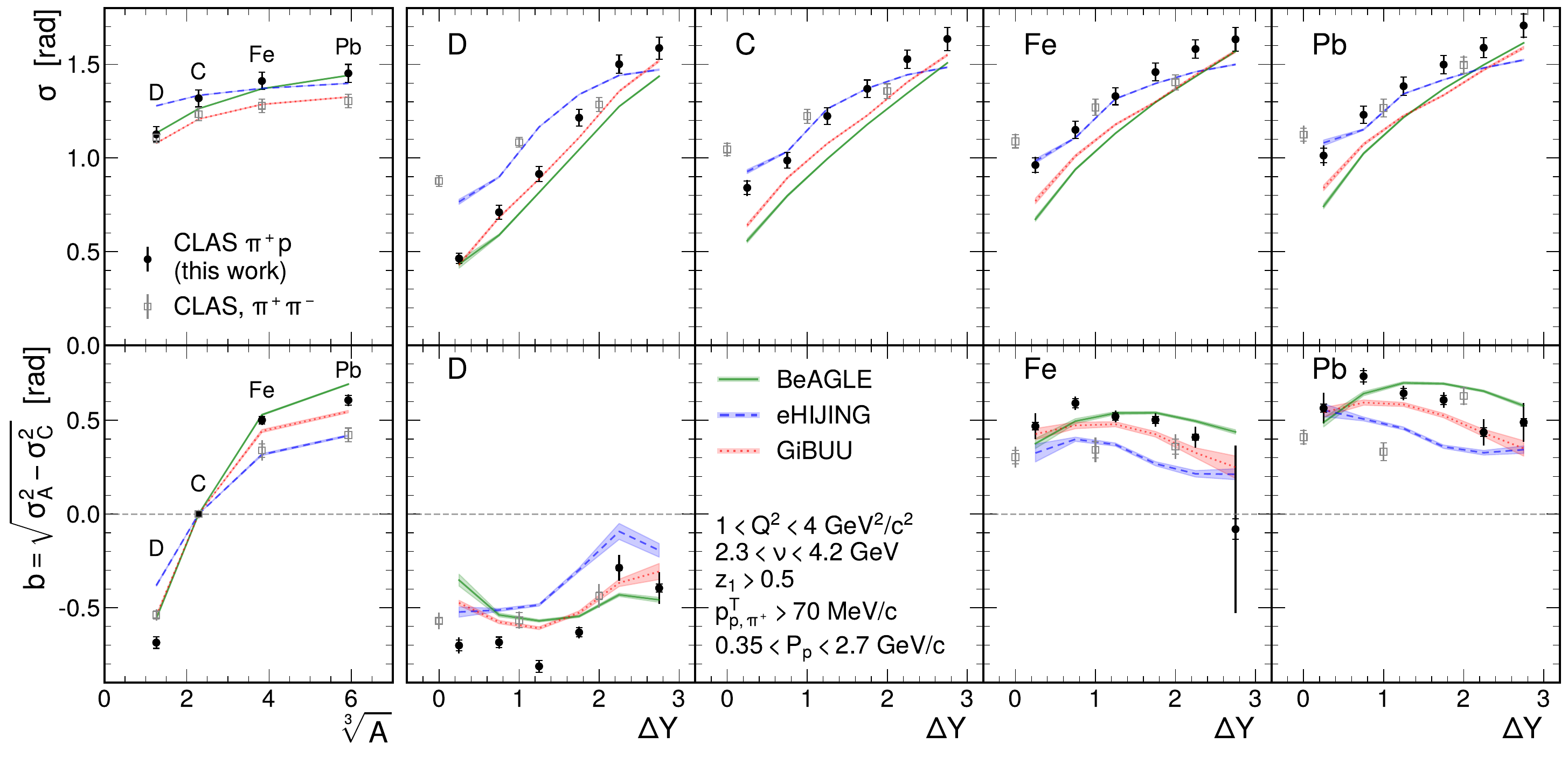}
    \caption{Top row: RMS width $\sigma$ as a function of $\sqrt[3]A$ (left-most panel), and as a function of $\Delta Y$ for each target (other panels).
    Bottom row: same for the azimuthal broadening,  $b$. The horizontal caps in the error bars represent the systematic uncertainties, while the outer bars represent the total systematic and statistical uncertainty (added in quadrature).
    The results are compared to the predictions from the \textsc{BeAGLE}~\cite{Chang:2022hkt}, eHIJING~\cite{PhysRevD.110.034001}, and \textsc{GiBUU}~\cite{Buss:2011mx} models, as well as analogous measurements of the dipion channel from Ref.~\cite{PhysRevC.111.035201} (gray, open squares).  The broadenings for the dipion channel were reported in Ref.~\cite{PhysRevC.111.035201} with respect to deuterium, and have been re-calculated with respect to carbon in order to compare them to the results of this work.  
    }
    \label{fig:derived}
\end{figure*}

From the correlation functions, we determined the RMS widths, $\sigma$, integrated in $\Delta Y$, which we show in the top left panel of Fig.~\ref{fig:derived} as a function of $\sqrt[3]{A}$ (which proxies the radius of the nucleus).  We note that $\sigma$ increases monotonically with respect to increasing nuclear size, with values of about 1.13, 1.32, 1.41, and 1.45 for D, C, Fe, and Pb, respectively.
This increase is further reflected in the azimuthal broadenings, $b$, which increase as a function of $A$, as shown in the lower left panel of the same figure.
This trend is similar to what was observed in the dipion measurements~\cite{PhysRevC.111.035201} (gray open squares in Fig.~\ref{fig:derived}).  We note that this increase in width with respect to nuclear size is steeper in this work than in Ref.~\cite{PhysRevC.111.035201}. 
 This seems to suggest that low-energy protons undergo stronger final-state interactions than secondary pions.   In the context of some models, such as the \textsc{GiBUU} model~\cite{Buss:2011mx}, when secondary pions are produced, they have a ``pre-hadronic'' phase when they are color neutral but are not fully formed.  During this phase, the ``pre-hadrons'' have a weaker interaction with the nuclear material than after they have fully formed.  Since protons from knockout are already fully formed, they do not have this pre-hadronic phase; therefore they interact with the nucleus as soon as they are struck by the cascade. Another possible explanation is that the protons have lower kinetic energy than pions of the same momentum, and are therefore more susceptible to energy loss and absorption effects.

In the subsequent columns of this figure, we show the widths and broadenings as a function of $\Delta Y$ for each target.  The widths increase monotonically as a function of $\Delta Y$ (which was also observed for the dipion channel in Ref.~\cite{PhysRevC.111.035201}, however we observe that the slope of this dependence is larger for the $\pi p$ channel than for the dipion channel).  The broadening is consistently negative (up to $\approx -$0.8) for deuterium and positive for iron (up to $\approx 0.6$) and lead (up to $\approx 0.7$), with the exception of one bin in which the uncertainties are very large.    

These trends in the widths and broadenings are reproduced by the \textsc{BeAGLE} model, which predicts the values of $\sigma$ within about one error bar for the integrated results (upper left panel of Fig.~\ref{fig:derived}).  However, the agreement with the data is not as good within slices in $\Delta Y$ as in the integrated results.  This appears to be related to the discrepancy mentioned above, where the predicted correlation functions appear to be shifted in $\Delta Y$ relative to the data.

We also compared our results for the widths and broadenings to the predictions of the eHIJING model~\cite{PhysRevD.110.034001} (version 1.0) and the \textsc{GiBUU} model~\cite{Buss:2011mx} (2021 release) in Fig.~\ref{fig:derived}.  We also provide a comparison between these models' predictions for the correlation functions and the ratio, $R$, in the Supplementary Materials~\cite{supplementary}. 
In \eHIJING, the hard-scattering process, along with initial- and final-state fragmentation, is simulated using \textsc{PYTHIA8}~\cite{bierlich2022comprehensive}. The model incorporates nuclear modifications to PDFs, fragmentation, and parton showering.  It also handles multiple collisions between partons and the rest of the nucleus, with a cross section proportional to the transverse-momentum-dependent gluon distribution density.  Like the BeAGLE model, eHIJING was developed for the Electron-Ion Collider.  

The \textsc{GiBUU} model incorporates treatment of final-state interactions, absorption, and production mechanisms with elastic and inelastic channels. The \textsc{GiBUU} model describes the single-hadron and di-hadron data reasonably well from CLAS~\cite{CLAS:2021jhm,CLAS:2022asf,PhysRevC.111.035201} and HERMES~\cite{HERMES:2003icw,Airapetian:2007vu,Airapetian:2011jp}.   We modified the default values of several parameters related  to parton distributions and fragmentation in free nucleons, based on a tuning study
that used an independent proton-target dataset from Ref.~\cite{CLAS:2022sqt}, and which was further validated with dipion correlations in Ref.~\cite{PhysRevC.111.035201}.  This tuning does not modify the nuclear effects.

 Both the eHIJING and \textsc{GiBUU} models reproduce the trends of the data.  The eHIJING predictions have a smaller variation of $\sigma$ with respect to both nuclear size and $\Delta Y$.  Further, it underestimates the magnitude of the broadening compared to the data.  The GiBUU model predicts the widths very well for deuterium (both in the $\Delta Y$ bins and when integrated in $\Delta Y$), but underpredicts the widths for the nuclear targets.   

\section{Summary and conclusions} 
We have reported the first measurement of angular correlations between leading pions and protons emerging from nuclear fragmentation, opening a new avenue to study how nuclei respond to fast hadrons.

Using electron–nucleus scattering data collected with the CLAS detector on deuterium and a broad set of heavier targets, we measured correlation functions that peak when the two particles are azimuthally back‑to‑back and their rapidity difference matches that between the virtual\hyp{}photon+struck-nucleon frame and the initial‑nucleus rest frame. 
As the rapidity separation increases, the azimuthal peak broadens; this broadening is more pronounced in larger nuclei.

The azimuthal‑peak width grows systematically with nuclear size, and the proton yield per pion also rises but it appears to saturate for the heaviest targets. These patterns are consistent with a picture in which hadron absorption and interactions in the hadronic cascade widen the correlation and the larger pool of protons in heavier nuclei enhances the yield. 

Compared with earlier dipion measurements~\cite{PhysRevC.111.035201}, the $\pi p$ channel shows a markedly stronger dependence on both nuclear size and rapidity separation. The trends are qualitatively reproduced by the state-of-the-art $eA$ event-generator models BeAGLE, eHIJING, and \textsc{GiBUU}, but quantitative differences remain, highlighting opportunities to refine the treatment of nuclear fragmentation.

Our results establish a benchmark for forthcoming di-hadron and slow-nucleon production studies in DIS, as well as for the development of corresponding event-generator models for experiments at Jefferson Lab~\cite{Burkert:2018nvj,Burkert:2020akg,Arrington:2021alx} and at future Electron–Ion Colliders~\cite{Accardi:2012qut,Anderle:2021wcy}.

\section{Code and Data Availability}

We provide in Ref.~\cite{zenodoRivet} the code we used to extract the $(\Delta\phi,\Delta Y)$ histograms from the output files of event generators (in order to produce the correlation functions) in the form of a module in the Rivet framework~\cite{10.21468/SciPostPhys.8.2.026}.

The values of the correlation functions, the ratio $R$, RMS widths, and broadenings reported in this work can be found in the Supplementary Materials~\cite{supplementary}.  
\section{Acknowledgements} 
The authors acknowledge the staff of the Accelerator and Physics Divisions at 
the Thomas Jefferson National Accelerator Facility who made this experiment 
possible.
We thank Kai Gallmeister for help in setting up the \textsc{GiBUU} event generator and Weiyao Ke for help in setting up the eHIJING event generator. This work is supported by the Chilean Agencia Nacional de Investigaci\'on y Desarrollo (ANID), FONDECYT Grants No.~1221827 and No.~1240904, ANID PIA/APOYO AFB230003 and by the ANID-Millennium Science Initiative Program -- ICN2019\_044, by the U.S.
Department of Energy, the Italian Instituto Nazionale di Fisica Nucleare, the French Centre 
National de la Recherche Scientifique, the French Commissariat \`a l'Energie 
Atomique, 
the United Kingdom Science and Technology Facilities Council (STFC), the 
Scottish Universities Physics Alliance (SUPA), the National Research Foundation 
of Korea, the National Science Foundation (NSF),  the HelmholtzForschungsakademie Hessen für FAIR (HFHF), the Ministry of Science and Higher Education of the Russian Federation, and the Office of Research and Economic Development at Mississippi 
State University. This work has received funding from 
the European Research Council (ERC) under the European Union’s Horizon 2020 
research and innovation programme (Grant agreement No.~804480). The Southeastern Universities Research 
Association operates the Thomas Jefferson National Accelerator Facility for the 
United States Department of Energy under Contract No.~DE-AC05-06OR23177.

 \FloatBarrier
\bibliographystyle{apsrev4-1}
\bibliography{biblio.bib} % refers to example.bib

@article{CLAS:2022asf,
    author = "Paul, S. J. and others",
    collaboration = "\textit{CLAS Collaboration}",
    title = "{Observation of Azimuth-Dependent Suppression of Hadron Pairs in Electron Scattering off Nuclei}",
    _eprint = "2207.06682",
    _archivePrefix = "arXiv",
    primaryClass = "nucl-ex",
    doi = "10.1103/PhysRevLett.129.182501",
    journal = "Phys. Rev. Lett.",
    volume = "129",
    number = "18",
    pages = "182501",
    year = "2022"
}

@article{CLAS:2021jhm,
    author = "Moran, S. and others",
    collaboration = "\textit{CLAS Collaboration}",
    title = "{Measurement of charged-pion production in deep-inelastic scattering off nuclei with the CLAS detector}",
    _eprint = "2109.09951",
    _archivePrefix = "arXiv",
    primaryClass = "nucl-ex",
    doi = "10.1103/PhysRevC.105.015201",
    journal = "Phys. Rev. C",
    volume = "105",
    number = "1",
    pages = "015201",
    year = "2022"
}

@article{CLAS:2022oux,
    author = "Chetry, T. and others",
    collaboration = "\textit{CLAS Collaboration}",
    title = "{First Measurement of \ensuremath{\Lambda} Electroproduction off Nuclei in the Current and Target Fragmentation Regions}",
    _eprint = "2210.13691",
    _archivePrefix = "arXiv",
    primaryClass = "nucl-ex",
    reportNumber = "JLAB-PHY-22-3746",
    doi = "10.1103/PhysRevLett.130.142301",
    journal = "Phys. Rev. Lett.",
    volume = "130",
    number = "14",
    pages = "142301",
    year = "2023"
}

@article{E665:1994aiu,
    author = "Adams, M. R. and others",
    collaboration = "E665",
    title = "{Nuclear Shadowing, Diffractive Scattering and Low Momentum Protons in $\mu Xe$ Interactions at 490 GeV}",
    reportNumber = "FERMILAB-PUB-94-218-E",
    doi = "10.1007/BF01571879",
    journal = "Z. Phys. C",
    volume = "65",
    pages = "225--244",
    year = "1995"
}

@Article{10.21468/SciPostPhys.8.2.026,
	title={{Robust Independent Validation of Experiment and Theory: Rivet version 3}},
	author={Christian Bierlich and Andy Buckley and Jonathan Butterworth and Christian Holm Christensen and Louie Corpe and David Grellscheid and Jan Fiete Grosse-Oetringhaus and Christian Gutschow and Przemyslaw Karczmarczyk and Jochen Klein and Leif Lonnblad and Christopher Samuel Pollard and Peter Richardson and Holger Schulz and Frank Siegert},
	journal={SciPost Phys.},
	volume={8},
	pages={026},
	year={2020},
	publisher={SciPost},
	doi={10.21468/SciPostPhys.8.2.026},
	url={https://scipost.org/10.21468/SciPostPhys.8.2.026},
}

@article{Buss:2011mx,
    author = "Buss, O. and Gaitanos, T. and Gallmeister, K. and van Hees, H. and Kaskulov, M. and Lalakulich, O. and Larionov, A. B. and Leitner, T. and Weil, J. and Mosel, U.",
    title = "{Transport-theoretical Description of Nuclear Reactions}",
    _eprint = "1106.1344",
    _archivePrefix = "arXiv",
    primaryClass = "hep-ph",
    doi = "10.1016/j.physrep.2011.12.001",
    journal = "Phys. Rept.",
    volume = "512",
    pages = "1--124",
    year = "2012"
}

@misc{zenodoRivet,
    author="Sebouh J. Paul",
    title="{Rivet Module for Pion-Proton Correlations in CLAS data}",
    doi="10.5281/zenodo.15492483",
}

@article{Ferrari:2005zk,
    author = "Ferrari, Alfredo and Sala, Paola R. and Fasso, Alberto and Ranft, Johannes",
    title = "{FLUKA: A Multi-Particle Transport Code}",
    reportNumber = "CERN-2005-010, SLAC-R-773, INFN-TC-05-11, CERN-2005-10",
    doi = "10.2172/877507",
    month = "10",
    year = "2005",
    note="(Program version 2005)",
    url="https://doi.org/10.2172/877507"
}

@misc{supplementary,
note = "See Supplemental Material for tables of the correlation functions, ratio $R$, RMS widths, and broadenings, as well as details on the systematic uncertainties and corrections to the data."
}

@inproceedings{Roesler:2000he,
    author = "Roesler, Stefan and Engel, Ralph and Ranft, Johannes",
    title = "{The Monte Carlo Event Generator DPMJET-III}",
    booktitle = "{International Conference on Advanced Monte Carlo for Radiation Physics, Particle Transport Simulation and Applications (MC 2000)}",
    reportNumber = "SLAC-PUB-8740",
    doi = "10.1007/978-3-642-18211-2_166",
    pages = "1033--1038",
    month = "12",
    year = "2000"
}

@article{Hen:2014nza,
    author = "Hen, O. and others",
    collaboration="\textit{CLAS Collaboration}",
    title = "{Momentum sharing in imbalanced Fermi systems}",
    _eprint = "1412.0138",
    _archivePrefix = "arXiv",
    primaryClass = "nucl-ex",
    doi = "10.1126/science.1256785",
    journal = "Science",
    volume = "346",
    pages = "614--617",
    year = "2014"
}

@article{CLAS:2022sqt,
    author = "Avakian, H. and others",
    collaboration = "\textit{CLAS Collaboration}",
    title = "{Observation of Correlations between Spin and Transverse Momenta in Back-to-Back Dihadron Production at CLAS12}",
    _eprint = "2208.05086",
    _archivePrefix = "arXiv",
    primaryClass = "hep-ex",
    reportNumber = "JLAB-PHY-22-3690",
    doi = "10.1103/PhysRevLett.130.022501",
    journal = "Phys. Rev. Lett.",
    volume = "130",
    number = "2",
    pages = "022501",
    year = "2023"
}

@article{CLAS:2003umf,
    author = "Mecking, B. A. and others",
    title = "{The CEBAF Large Acceptance Spectrometer (CLAS)}",
    reportNumber = "JLAB-PHY-03-01",
    doi = "10.1016/S0168-9002(03)01001-5",
    journal = "Nucl. Instrum. Meth. A",
    volume = "503",
    pages = "513--553",
    year = "2003"
}

@phdthesis{Hakobyan:2008kua,
    author = "Hakobyan, Hayk",
    title = "{Observation of Quark Propagation Pattern in Nuclear Medium}",
    school = "Yerevan State U.",
    year = "2008",
    url="https://www.jlab.org/Hall-B/general/thesis/Hakobyan_thesis.pdf"
}

@misc{GSIM,
      author         = "Wolin, E",
      title          = "{GSIM User’s Guide Version 1.1}",
      year           = "1996",
      url            = "https://www.jlab.org/Hall-B/document/gsim/userguide.html"
}

@report{LRP2023,
    author="{Nuclear Science Advisory Committee}",
    title="{A New Era of Discovery: The 2023 Long Range Plan for Nuclear Science}",
    note="\url{https://nuclearsciencefuture.org/wp-content/uploads/2023/11/NSAC-LRP-2023-v1.3.pdf}"
}

@article{Brun:1994aa,
      author         = "Brun, René and Bruyant, F. and Carminati, Federico and
                        Giani, Simone and Maire, M. and McPherson, A. and Patrick,
                        G. and Urban, L.",
      title          = "{GEANT Detector Description and Simulation Tool}",
      doi            = "10.17181/CERN.MUHF.DMJ1",
      year           = "1994",
      reportNumber   = "CERN-W5013, CERN-W-5013, W5013, W-5013",
      SLACcitation   = "%%CITATION = CERN-W5013;%%"
}

@article{CLAS:2012tlh,
    author = "El Fassi, L. and others",
    collaboration = "\textit{CLAS Collaboration}",
    title = "{Evidence for the onset of color transparency in $\rho^0$  electroproduction off nuclei}",
    _eprint = "1201.2735",
    _archivePrefix = "arXiv",
    primaryClass = "nucl-ex",
    reportNumber = "JLAB-PHY-12-1504",
    doi = "10.1016/j.physletb.2012.05.019",
    journal = "Phys. Lett. B",
    volume = "712",
    pages = "326--330",
    year = "2012"
}

@article{GEANT4:2002zbu,
    author = "Agostinelli, S. and others",
    collaboration = "GEANT4",
    title = "{GEANT4 - A Simulation Toolkit}",
    reportNumber = "SLAC-PUB-9350, FERMILAB-PUB-03-339, CERN-IT-2002-003",
    doi = "10.1016/S0168-9002(03)01368-8",
    journal = "Nucl. Instrum. Meth. A",
    volume = "506",
    pages = "250--303",
    year = "2003"
}

@article{HERMES:2005mar,
    author = "Airapetian, A. and others",
    collaboration = "\textit{HERMES Collaboration}",
    title = "{Double hadron leptoproduction in the nuclear medium}",
    _eprint = "hep-ex/0510030",
    _archivePrefix = "arXiv",
    reportNumber = "DESY-05-205",
    doi = "10.1103/PhysRevLett.96.162301",
    journal = "Phys. Rev. Lett.",
    volume = "96",
    pages = "162301",
    year = "2006"
}

@article{Fialkowski:2007pt,
    author = "Fialkowski, K. and Wit, R.",
    title = "{On the electroproduction on nuclei}",
    _eprint = "hep-ph/0702058",
    _archivePrefix = "arXiv",
    doi = "10.1140/epja/i2007-10361-2",
    journal = "Eur. Phys. J. A",
    volume = "32",
    pages = "213--218",
    year = "2007"
}

@article{Airapetian:2007vu,
      author         = "Airapetian, A. and others",
      title          = "{Hadronization in semi-inclusive deep-inelastic
                        scattering on nuclei}",
      collaboration  = "\textit{HERMES Collaboration}",
      journal        = "Nucl. Phys.",
      volume         = "B780",
      year           = "2007",
      pages          = "1-27",
      doi            = "10.1016/j.nuclphysb.2007.06.004",
      _eprint         = "0704.3270",
      archivePrefix  = "arXiv",
      primaryClass   = "hep-ex",
      reportNumber   = "DESY-07-050",
      SLACcitation   = "%%CITATION = ARXIV:0704.3270;%%"
}

@article{Airapetian:2011jp,
      author         = "Airapetian, A. and others",
      title          = "{Multidimensional Study of Hadronization in Nuclei}",
      collaboration  = "\textit{HERMES Collaboration}",
      journal        = "Eur. Phys. J.",
      volume         = "A47",
      year           = "2011",
      pages          = "113",
      doi            = "10.1140/epja/i2011-11113-5",
      _eprint         = "1107.3496",
      archivePrefix  = "arXiv",
      primaryClass   = "hep-ex",
      reportNumber   = "DESY-11-120",
      SLACcitation   = "%%CITATION = ARXIV:1107.3496;%%"
}

@article{Anderle:2021wcy,
    author = "Anderle, Daniele P. and others",
    title = "{Electron-ion collider in China}",
    _eprint = "2102.09222",
    archivePrefix = "arXiv",
    primaryClass = "nucl-ex",
    reportNumber = "Frontiers of Physics, Volume 16 Issue (6):64701, 2021",
    doi = "10.1007/s11467-021-1062-0",
    journal = "Front. Phys. (Beijing)",
    volume = "16",
    number = "6",
    pages = "64701",
    year = "2021"
}

@article{Accardi:2012qut,
    author = "Accardi, A. and others",
    editor = "Deshpande, A. and Meziani, Z. E. and Qiu, J. W.",
    title = "{Electron Ion Collider: The Next QCD Frontier}: {Understanding the glue that binds us all}",
    _eprint = "1212.1701",
    _archivePrefix = "arXiv",
    primaryClass = "nucl-ex",
    reportNumber = "BNL-98815-2012-JA, JLAB-PHY-12-1652",
    doi = "10.1140/epja/i2016-16268-9",
    journal = "Eur. Phys. J. A",
    volume = "52",
    number = "9",
    pages = "268",
    year = "2016"
}

@article{AbdulKhalek:2021gbh,
    author = "Abdul Khalek, R. and others",
    title = "{Science Requirements and Detector Concepts for the Electron-Ion Collider}: {EIC Yellow Report}",
    _eprint = "2103.05419",
    _archivePrefix = "arXiv",
    primaryClass = "physics.ins-det",
    reportNumber = "BNL-220990-2021-FORE, JLAB-PHY-21-3198, LA-UR-21-20953",
    doi = "10.1016/j.nuclphysa.2022.122447",
    journal = "Nucl. Phys. A",
    volume = "1026",
    pages = "122447",
    year = "2022"
}

@article{Burkert:2018nvj,
    author = "Burkert, Volker D.",
    title = "{Jefferson Lab at 12 GeV: The Science Program}",
    doi = "10.1146/annurev-nucl-101917-021129",
    journal = "Ann. Rev. Nucl. Part. Sci.",
    volume = "68",
    pages = "405--428",
    year = "2018"
}

@article{Burkert:2020akg,
    author = "Burkert, V. D. and others",
    title = "{The CLAS12 Spectrometer at Jefferson Laboratory}",
    collaboration="\textit{CLAS Collaboration}",
    doi = "10.1016/j.nima.2020.163419",
    journal = "Nucl. Instrum. Meth. A",
    volume = "959",
    pages = "163419",
    year = "2020"
}

@article{Arrington:2021alx,
    author = "Arrington, J. and others",
    title = "{Physics with CEBAF at 12 GeV and future opportunities}",
    _eprint = "2112.00060",
    _archivePrefix = "arXiv",
    primaryClass = "nucl-ex",
    doi = "10.1016/j.ppnp.2022.103985",
    journal = "Prog. Part. Nucl. Phys.",
    volume = "127",
    pages = "103985",
    year = "2022"
}

@article{PhysRevD.100.073010,
  title = {Pion-proton correlation in neutrino interactions on nuclei},
  author = {Cai, Tejin and Lu, Xianguo and Ruterbories, Daniel},
  journal = {Phys. Rev. D},
  volume = {100},
  issue = {7},
  pages = {073010},
  numpages = {7},
  year = {2019},
  month = {Oct},
  publisher = {American Physical Society},
  doi = {10.1103/PhysRevD.100.073010},
  url = {https://link.aps.org/doi/10.1103/PhysRevD.100.073010}
}

@article{Degtyarenko:1997,
    author=
    "Degtyarenko, P.V. and Doroshkevich, E.A. and  Kurzenkov and A.A. and others",
    title="Proton-pion and two- pion correlations in $eA$ interactions.",
    journal="Z Phys A - Particles and Fields",
    volume="357", 
    pages="419–424",
    year="1997",
    doi={10.1007/s002180050261}
}

@article{Palli:2009it,
    author = "Palli, V. and Ciofi degli Atti, C. and Kaptari, L. P. and Mezzetti, C. B. and Alvioli, M.",
    title = "{Slow Proton Production in Semi-Inclusive Deep Inelastic Scattering off Deuteron and Complex Nuclei: Hadronization and Final State Interaction Effects}",
    _eprint = "0911.1377",
    _archivePrefix = "arXiv",
    primaryClass = "nucl-th",
    doi = "10.1103/PhysRevC.80.054610",
    journal = "Phys. Rev. C",
    volume = "80",
    pages = "054610",
    year = "2009"
}

@article{E910:1999ozb,
    author = "Chemakin, I. and others",
    collaboration = "\textit{E910 Collaboration}",
    title = "{Measuring centrality with slow protons in proton nucleus collisions at the AGS}",
    _eprint = "nucl-ex/9902003",
    archivePrefix = "arXiv",
    doi = "10.1103/PhysRevC.60.024902",
    journal = "Phys. Rev. C",
    volume = "60",
    pages = "024902",
    year = "1999"
}

@article{PhysRevC.111.035201,
  title = {Dihadron azimuthal correlations in deep-inelastic scattering off nuclear targets},
  author = {Paul, S. J. and others},
  collaboration = "\textit{CLAS Collaboration}",
  journal = {Phys. Rev. C},
  volume = {111},
  issue = {3},
  pages = {035201},
  numpages = {24},
  year = {2025},
  month = {Mar},
  publisher = {American Physical Society},
  doi = {10.1103/PhysRevC.111.035201},
  url = {https://link.aps.org/doi/10.1103/PhysRevC.111.035201}
}

@article{Chang:2022hkt,
    author = "Chang, Wan and Aschenauer, Elke-Caroline and Baker, Mark D. and Jentsch, Alexander and Lee, Jeong-Hun and Tu, Zhoudunming and Yin, Zhongbao and Zheng, Liang",
    title = "{Benchmark $eA$ generator for leptoproduction in high-energy lepton-nucleus collisions}",
    _eprint = "2204.11998",
    archivePrefix = "arXiv",
    primaryClass = "physics.comp-ph",
    doi = "10.1103/PhysRevD.106.012007",
    journal = "Phys. Rev. D",
    volume = "106",
    number = "1",
    pages = "012007",
    year = "2022"
}

@inproceedings{Whalley:2005nh,
    author = "Whalley, M. R. and Bourilkov, D. and Group, R. C.",
    title = "{The Les Houches accord PDFs (LHAPDF) and LHAGLUE}",
    booktitle = "{HERA and the LHC: A Workshop on the Implications of HERA and LHC Physics (Startup Meeting, CERN, 26-27 March 2004; Midterm Meeting, CERN, 11-13 October 2004)}",
    pages = "575--581",
    month = "8",
    year = "2005",
    doi = "10.5170/CERN-2005-014.575"
}

@article{kbhz-h4jv,
  title = {Suppression of neutral-pion production in deep-inelastic scattering off nuclei with the CLAS detector},
  author = {Mineeva, T. and Brooks, W. K. and El Alaoui, A. and Hakobyan, H. and Joo, K. and L\'opez, J. A. and Soto, O. and others},
  collaboration = {\textit{CLAS Collaboration}},
  journal = {Phys. Rev. C},
  volume = {112},
  issue = {3},
  pages = {035203},
  numpages = {14},
  year = {2025},
  month = {Sep},
  publisher = {American Physical Society},
  doi = {10.1103/kbhz-h4jv},
  url = {https://link.aps.org/doi/10.1103/kbhz-h4jv}
}

@article{bierlich2022comprehensive,
author = {Bierlich, Christian and Chakraborty, Smita and Desai, Nishita and Gellersen, Leif and Helenius, Ilkka and Ilten, Philip and Lönnblad, Leif and Mrenna, Stephen and Prestel, Stefan and Preuss, Christian and Sjöstrand, Torbjörn and Skands, Peter and Utheim, Marius and Verheyen, Rob},
year = {2022},
month = {11},
pages = {},
title = {A comprehensive guide to the physics and usage of PYTHIA 8.3},
journal = {SciPost Physics Codebases},
doi = {10.21468/SciPostPhysCodeb.8}
}

@article{Sjostrand:2006za,
    author = "Sjostrand, Torbjorn and Mrenna, Stephen and Skands, Peter Z.",
    title = "{PYTHIA 6.4 Physics and Manual}",
    _eprint = "hep-ph/0603175",
    archivePrefix = "arXiv",
    reportNumber = "FERMILAB-PUB-06-052-CD-T, LU-TP-06-13",
    doi = "10.1088/1126-6708/2006/05/026",
    journal = "JHEP",
    volume = "05",
    pages = "026",
    year = "2006"
}

@phdthesis{Dupre:2011afa,
    author = {Dupr\'e, Rapha\"el},
    title = "{Quark Fragmentation and Hadron Formation in Nuclear Matter}",
    reportNumber = "tel-00670293, tel-00751424",
    school = "Lyon, IPN",
    year = "2011",
    url="https://theses.hal.science/tel-00751424/document"
}

@article{RoblesGajardo:2022efe,
    author = {Robles Gajardo, Carolina M. and Accardi, Alberto and Baker, Mark D. and Brooks, William K. and Dupr\'e, Rapha\"el and Ehrhart, Mathieu and L\'opez, Jorge A. and Tu, Zhoudunming},
    title = "{Low energy protons as probes of hadronization dynamics}",
    _eprint = "2203.16665",
    _archivePrefix = "arXiv",
    primaryClass = "hep-ph",
    doi = "10.1103/PhysRevC.106.045202",
    journal = "Phys. Rev. C",
    volume = "106",
    number = "4",
    pages = "045202",
    year = "2022"
}

@article{Larionov:2018igy,
    author = "Larionov, A. B. and Strikman, M.",
    title = "{Slow neutron production as a probe of hadron formation in high-energy $\gamma^*A$ reactions}",
    _eprint = "1812.08231",
    archivePrefix = "arXiv",
    primaryClass = "hep-ph",
    doi = "10.1103/PhysRevC.101.014617",
    journal = "Phys. Rev. C",
    volume = "101",
    number = "1",
    pages = "014617",
    year = "2020"
}

@article{HERMES:2003icw,
    author = "Airapetian, A. and others",
    collaboration = "\textit{HERMES Collaboration}",
    title = "{Quark fragmentation to $\pi^\pm$, $\pi^0$, $K^\pm$, $p$ and $\bar p$ in the nuclear environment}",
    _eprint = "hep-ex/0307023",
    archivePrefix = "arXiv",
    reportNumber = "DESY-03-088",
    doi = "10.1016/j.physletb.2003.10.026",
    journal = "Phys. Lett. B",
    volume = "577",
    pages = "37--46",
    year = "2003"
}

@article{Airapetian:2009jy,
      author         = "Airapetian, A. and others",
      title          = "{Transverse momentum broadening of hadrons produced in
                        semi-inclusive deep-inelastic scattering on nuclei}",
      collaboration =  "\textit{HERMES Collaboration}",
      journal        = "Phys.~Lett.~B",
      volume         = "684",
      year           = "2010",
      pages          = "114-118",
      doi            = "10.1016/j.physletb.2010.01.020",
      _eprint = "arXiv:0906.2478",
      archivePrefix  = "arXiv",
      primaryClass   = "hep-ex",
      reportNumber   = "DESY-09-082",
      SLACcitation   = "%%CITATION = ARXIV:0906.2478;%%"
}

@article{HERMES:2000ytc,
    author = "Airapetian, A. and others",
    collaboration = "\textit{HERMES Collaboration}",
    title = "{Hadron formation in deep inelastic positron scattering in a nuclear environment}",
    _eprint = "hep-ex/0012049",
    archivePrefix = "arXiv",
    reportNumber = "DESY-00-191",
    doi = "10.1007/s100520100697",
    journal = "Eur. Phys. J. C",
    volume = "20",
    pages = "479--486",
    year = "2001"
}

@article{PhysRevD.110.034001,
  title = {Event generator for jet tomography in electron-ion collisions},
  author = {Ke, Weiyao and Zhang, Yuan-Yuan and Xing, Hongxi and Wang, Xin-Nian},
  journal = {Phys. Rev. D},
  volume = {110},
  issue = {3},
  pages = {034001},
  numpages = {31},
  year = {2024},
  month = {Aug},
  publisher = {American Physical Society},
  doi = {10.1103/PhysRevD.110.034001},
  url = {https://link.aps.org/doi/10.1103/PhysRevD.110.034001}
}

@article{BEBCWA59:1989ayi,
    author = "Matsinos, E. and others",
    collaboration = "\textit{BEBC WA59 Collaboration}",
    title = "{Backward Particle Production in Neutrino Neon Interactions}",
    reportNumber = "CERN-EP-89-57",
    doi = "10.1007/BF01548585",
    journal = "Z. Phys. C",
    volume = "44",
    pages = "79",
    year = "1989"
}

@article{CiofidegliAtti:2004pv,
    author = "Ciofi degli Atti, C. and Kopeliovich, B. Z.",
    title = "{Time evolution of hadronization and grey tracks in DIS off nuclei}",
    _eprint = "hep-ph/0409077",
    _archivePrefix = "arXiv",
    doi = "10.1016/j.physletb.2004.12.021",
    journal = "Phys. Lett. B",
    volume = "606",
    pages = "281--287",
    year = "2005"
}

@article{ALIAGA2014130,
title = {Design, calibration, and performance of the MINERvA detector},
journal = {Nuclear Instruments and Methods in Physics Research Section A: Accelerators, Spectrometers, Detectors and Associated Equipment},
volume = {743},
pages = {130-159},
year = {2014},
issn = {0168-9002},
doi = {https://doi.org/10.1016/j.nima.2013.12.053},
url = {https://www.sciencedirect.com/science/article/pii/S0168900214000035},
author = {L. Aliaga and others},
collaboration={\textit{MINERvA Collaboration}}
}

@techreport{osti_935497,
  author       = {Ayres, D S and Drake, G R and Goodman, M C and Grudzinski, J J and Guarino, V J and Talaga, R L and Zhao, A and Stamoulis, P and Stiliaris, E and Tzanakos, G and others},
  title        = {The NOvA Technical Design Report},
  institution  = {Fermi National Accelerator Laboratory (FNAL), Batavia, IL},
  doi          = {10.2172/935497},
  url          = {https://www.osti.gov/biblio/935497},
  place        = {United States},
  year         = {2007},
  month        = {10}}

\end{document}